\documentclass[12pt, a4paper, abstract=true, titlepage=false, parskip=half, headinclude]{scrartcl}
\pdfoutput=1

\usepackage[a4paper,nomarginpar,left=2.5cm,right=2.5cm,top=3cm,bottom=3.5cm]{geometry}

\usepackage{amsmath}

\usepackage{newclude}

\usepackage{fancyhdr}
\usepackage[dvipsnames]{xcolor}

\usepackage[colorlinks, citecolor=ForestGreen, urlcolor=MidnightBlue, linkcolor=RoyalBlue]{hyperref}
\usepackage[capitalise,noabbrev,nameinlink,sort]{cleveref}

\usepackage[sfdefault]{carlito}

\usepackage{csquotes}
\usepackage{float}
\usepackage{tcolorbox}
\tcbuselibrary{skins}
\definecolor{tboxbg}{HTML}{EEFAFC}

\tcbset{%
every box/.style={%
    enhanced,
    colback=tboxbg,
    colframe=teal,
    arc=15pt,
    colbacktitle=tboxbg,
    coltitle=teal,
    fonttitle={\large\bfseries},
    toptitle=7pt,
    bottomtitle=4pt,
    righttitle=2mm,
    top=0mm,
    boxrule=2pt,
    titlerule=0pt,
    titlerule style={tboxbg},
    fontupper=\footnotesize,
},
every box on layer 2/.style={reset, every box, arc=5pt},
every float=\centering,
}
\newtcolorbox[%
    auto counter,
    crefname={Toolbox}{Toolboxes},
    Crefname={Toolbox}{Toolboxes}]%
{toolbox}[2][]{%
    title=Toolbox \thetcbcounter: #2, #1}
\newtcolorbox[]{innertoolbox}[2][]{%
    title=#2,
    #1,
    halign title=center,
    fontupper=\scriptsize,
    toptitle=5pt,
    bottomtitle=2pt,
    fonttitle={\normalsize\bfseries},
}
\usepackage{capt-of}
\tcbuselibrary{raster}

\renewcommand{\arraystretch}{1.5}

\usepackage{graphicx}
\usepackage[textfont={it},font=footnotesize]{caption}

\usepackage[sorting=none, style=numeric]{biblatex}
\usepackage[separate-uncertainty=true, exponent-product = \cdot, range-units=single]{siunitx}
\usepackage{pdfpages}

\renewcommand{\_}[1]{_\mathrm{#1}}

\RedeclareSectionCommand[
  beforeskip=-\baselineskip,
  afterskip=.3\baselineskip]{section}
\RedeclareSectionCommand[
  beforeskip=-.3\baselineskip,
  afterskip=.2\baselineskip]{subsection}
\RedeclareSectionCommand[
  beforeskip=-.1\baselineskip,
  afterskip=.05\baselineskip]{subsubsection}
\usepackage{enumitem}
\setlist{nosep}

\usepackage{authblk}

\title{Active Intracellular Mechanics: A Key to Cellular Function and Organization}

\author[1]{Mohammad~Amin~Eskandari}
\author[1]{Jannis~Fischer}
\author[1]{Noémie~Veyret}
\author[1]{Dorian~Marx}
\author[1,2,*]{Timo Betz}

\affil[1]{\footnotesize Third Institute of Physics - Biophysics \\
Georg-August-Universität Göttingen \\
Friedrich-Hund-Platz 1 \\
37077 Göttingen \\
Germany}

\affil[2]{\footnotesize Cluster of Excellence 'Multiscale Bioimaging: From Molecular Machines to Networks of Excitable Cells' (MBExC)\\
Georg-August-Universität Göttingen \\
Universitätsmedizin Göttingen \\
Robert-Koch-Str. 40 \\
37075 Göttingen \\
Germany}
\affil[*]{timo.betz@phys.uni-goettingen.de}

\begin{document}

\maketitle
\begin{abstract}
    \pdfoutput=1
    \noindent
    While mechanobiology has demonstrated that precise control over mechanical properties at the whole-cell level is crucial for many biological functions, comparatively little attention has been paid to the intracellular mechanical properties. Experimental tools have only recently become available to adequately measure the viscoelasticity and activity of the cytosol, revealing, revealing that the active, non-equilibrium nature of the intracellular environment must be carefully considered.
    To explore the interplay between active forces and viscoelastic properties, it is helpful to consider our current understanding of intracellular active mechanics. In this review, we aim not only to provide an intuitive and quantitative introduction to the relevant physical concepts, but also to offer an overview of the proteins that establish intracellular active mechanics, highlighting their spatial and temporal variation with a particular focus on the role of activity.
    Although we are only beginning to uncover the importance of intracellular active mechanics for cellular mechanisms, it is increasingly clear that these properties must be precisely regulated to ensure proper cellular function.
\end{abstract}
\thispagestyle{empty}

\clearpage
\setcounter{page}{1}
\pagestyle{fancy}

\section{The importance of intracellular mechanics to biological function}
\label{sec:introduction}
\pdfoutput=1

The past decades have seen a disruptive recognition that biological functions and behavior are often deeply connected to mechanical properties and forces, leading to the emergence of a new field in biology called mechanobiology \cite{phuyal2023mechanobiology, vogel2018unraveling, goodwin2021mechanics,pavin2021mechanobiology}. At the heart of mechanobiology's success in explaining many biological phenomena lies the seamless incorporation of physics-based material science and physics with cell and tissue biology \cite{alert2020physical, ladoux2017mechanobiology, doostmohammadi2022physics, julicher2018hydrodynamic, urbanska2024single}. While it is now widely accepted that mechanical properties such as stiffness, viscosity, and force generation are inherently important for proper cellular function \cite{moeendarbary2014cell}, much of our current understanding is based on mechanics and force generation at the whole-cell \cite{urbanska2024single} and tissue \cite{nelson2024mechanobiology, hofemeier2021global} scale. In fact, we only start to understand the implications of mechanobiology for functions inside the cell.

Only during the past decade, researchers have developed the experimental tools needed to study intracellular mechanical properties, as these need to literally feel inside the cells, thus overcoming the stiff cell cortex\cite{vos2024characterizing, ahmed2015active, rodriguez2013review}. This progress was accompanied by a deeper understanding of intracellular transport mechanisms, such as diffusion \cite{bressloff2013stochastic} and active particle motion \cite{guo2014probing, ahmed2018active,hurst2021intracellular, mogre2020getting}. This opens the door to connecting on one hand the intracellular mechanics to cellular organization, and on the other the effects of local viscoelastic properties and active forces for biochemical reactions.

Yet why should intracellular mechanical properties matter for biological function? Knowing that many cellular processes can be explained by biochemistry rather than physics, it may be tempting to downplay the relevance of intracellular mechanics in general. However, in the context of the complex intracellular environment, it turns out that even chemical reactions are strongly influenced by mechanics \cite{persson2020cellular, molines2022physical, llopis2024artificial}. This is because the rates of many biochemical reactions become diffusion-limited \cite{loverdo2008enhanced, grebenkov2023diffusion}, meaning that reaction networks are directly affected by viscosity and local crowding \cite{tan2013molecular}. In turn, active metabolic forces can enhance reaction rates by increaseing particle mobility far beyond simple Brownian motion \cite{ahmed2018active,caspi2000enhanced, zhao2017enhanced}.

Especially this active, metabolically driven increase in diffusion has drastic effects that are often not very well appreciated. The significance of this can be illustrated by the following, striking example. From a biological point of view, a high protein density is desired as these are the molecules generating biological functions. At a given point, the high molecular crowding leads to a drastic reduction of mobility and Brownian motion, known as a jamming transition. At this density, transport and even chemical reactions would freeze as reactants cannot meet. To overcome this, a cell could either increase its temperature or decrease the protein density. Both does not happen, still, cells are obviously alive, and the mobility maintains a high level. Since the energy measured to explain the spontaneous active diffusion of probe particles in cells is in the order of 10 times the thermal energy \cite{turlier2016equilibrium, ahmed2018active, hurst2021intracellular, muenker2023intracellular}, one may ask the simple question, what temperature is required to explain this mobility inside cells assume equilibrium thermodynamics. As an answer, one obtains the surprising value of \qtyrange{3000}{10000}{\degreeCelsius}---similar to the surface temperature of the Sun. This is a nice example that first gives the impression that biology is overcoming the limits provided by physics. Of course biology stays consistent with the laws of physics, but solves this riddle by simply driving intracellular diffusion by active forces instead of relying on purely Brownian motion, eventually leading to active diffusion \cite{ahmed2018active, brangwynne2009intracellular}. Another effect of this active metabolically driven diffusion is that mechanical heterogeneity inside the cell can provide an efficient way to organize intracellular space without relying solely on directed transport processes to maintain this organization \cite{benichou2010geometry, loverdo2008enhanced}.

The intrinsic connection between intracellular material properties and the diverse biological functions of cells is a very active research topic, full of mysteries currently being investigated by an interdisciplinary group of biologists, chemists, physicists, and mathematicians. Over the past ten years, the field has made significant strides forward. The primary aim of this review is to provide an overview of the current knowledge and methods used to quantify intracellular mechanics, and the different approaches to determine the relation between biological function, intracellular mechanical properties and active forces.

This goal is divided into two parts. In the main text, we aim to provide an intuitive understanding of relevant physical quantities such as mechanical moduli, elasticity, viscosity, and active energy without requiring the reader to follow the sometimes cumbersome physical definitions. The formal details of these concepts are covered in separate toolboxes to allow interested readers some deeper insights without interrupting the flow of the main text, while references point out resources to better understand the physics and mathematics required to appreciate the full potential of intracellular active mechanics. 

\begin{figure}[htb!]
    \centering
    \includegraphics[width=0.6\textwidth]{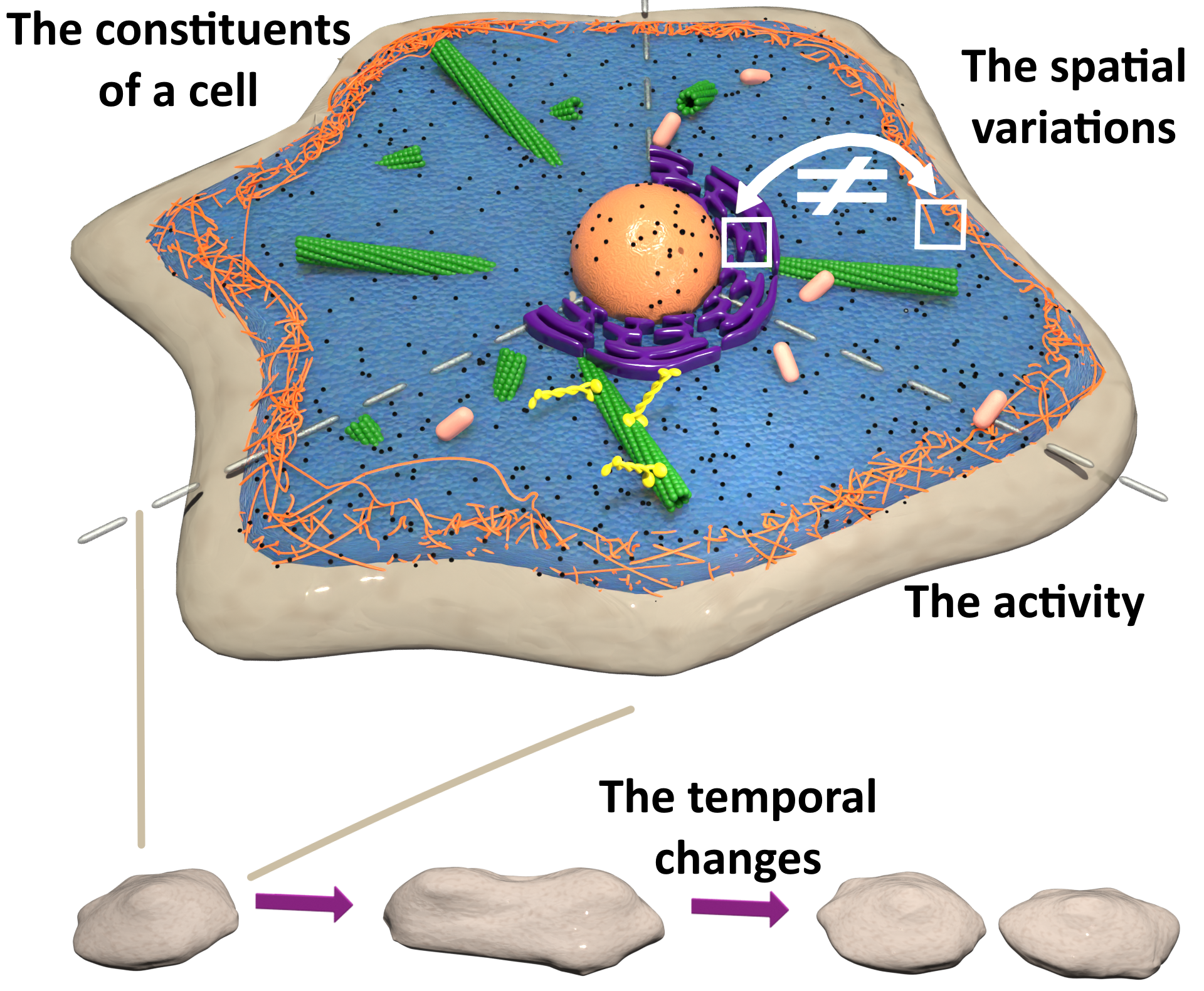}
    \caption{The review offers several different parts: After the Introduction to the topics and its theory and methods, \textbf{the constituents of a cell}, \textbf{the spatial variations} and  \textbf{the temporal changes} as well as \textbf{the activity} of the cytoplasm are discussed to cover the whole picture of intracellular mechanics.}
    \label{fig:IntroFigure}
\end{figure}

In addition to serving as an entry point to the field, this review discusses the knowledge of intracellular mechanics obtained over the past decades. We briefly introduce the main proteins known to establish and control intracellular active mechanics before focusing on spatial and temporal variations and the importance of active forces in determining mechanical properties and evaluating their implications for cell biology (\cref{fig:IntroFigure}).

\section{Getting intuition for viscoelasticity inside the cell}
\label{sec:theory}
\pdfoutput=1

Living in a physical world, we all possess a very good every-day understanding of fundamental concepts like elasticity, viscosity and activity as we continuously interact with many different kinds of materials around us. To harness this understanding, we will use examples from the world of food, namely sweets and deserts, to provide intuitive understanding of the mechanical concepts required to describe intracellular mechanics.In the following we will introduce examples of elastic, viscous and viscoelastic materials, starting with the concept of elasticity. A typical elastic material in this class of tasty examples is a gummy bear, or fruit gum. This collagen based material can be easily deformed when applying a force, but will return to its original shape after the force is released. The stiffness of such a material is commonly described by mechanical moduli. Reading about cell mechanics can be already confusing as some authors report the Young's modulus (for gummi bear: $E_\mathrm{G}\approx \qty{75}{\kilo\pascal}$), or a shear modulus ($G_\mathrm{G}\approx \qty{25}{\kilo\pascal}$) \cite{williams2005mechanical}. Both are related, but not the same property as the Young's modulus and the shear modulus quantify the resistance to stretching or shearing a material, respectively. For many materials the connection between Young's and Shear modulus is $E\approx 3 G$. The units of mechanical moduli are Pascal $[\unit{\pascal}] = [\unit{\newton}]/[\unit{\meter\squared}]$, which simply reflect that the extend of deformation as response to a force will depend on the area the force is applied to. As a rule of thumb, the modulus corresponds to the stress (force per area) that must be apply to stretch an object to twice its initial size (or to press to half its size). Again the fruit gum is an excellent example to get an idea about the stiffness of many biological tissues, since for example muscle or also the earlobe \cite{mckee2011indentation,vathulya2021comparison}, are all in a similar range as the fuit gum. In contrast, stiffness values relevant to the intracellular situations are quite soft, as these range from \qty{1}{\pascal} (cytosol) up to \qtyrange{1}{10}{\kilo\pascal} (cortex, nucleus). Everyday examples for such stiffnesses range from Ketchup (\qtyrange{10}{100}{\pascal}) over Mayonnaise (\qtyrange{100}{1000}{\pascal}) to whipped cream which can go up to \qty{10}{\kilo\pascal}.

In stark contrast to elastic solids are the viscous fluids. A fluid is in general a material that cannot sustain any shear forces and will immediately start flowing. The physical quantity that quantifies this fluid behaviour is the viscosity, which is measured in units of $\unit{\pascal}\times \unit{\second} \equiv \unit{\pascal\second}$. Since viscosity measures the resistance to flow, it is a time-dependent quantity and time is included in its units. Relevant viscosities for the context of intracellular mechanics range from \qty{0.003}{\pascal\second} up to \qty{0.5}{\pascal\second}. This corresponds to the viscosity of whole milk $\eta\_M \approx \qty{0.003}{\pascal\second}$, over heavy cream $\eta\_C \approx \qty{0.01}{\pascal\second}$ and up to  milkshakes $\eta\_{MS} \approx \qtyrange{0.05}{0.5}{\pascal\second}$. For reference, water at room temperature has a viscosity of about $\eta\_W \approx \qty{0.001}{\pascal\second}$.

Although it often seems that one can describe real materials by either their elastic modulus or viscosity, it turns out that almost all relevant materials are a mixture of both. As consequence, neither viscosity, nor elasticity are a good description for most of the materials. In fact, to describe a real viscoelastic material, the viscosity and also the elasticity become time dependent properties. This means that under a short-term force application, the material behaves differently under long-term force application. An example to illustrate this is a dessert like jelly. It stays elastic on short timescales, for example when poking it, but if one puts a spoon on it and waits for several hours, the spoon starts to sink in. This concept is very important for the description of the intracellular environment, as its properties depend drastically on timescales. In fact, often the material properties of the intracellular environment can be described by powerlaws (See \cref{tbox:mech_models} ), which are very useful to separate regimes where more fluid-like properties dominate (typically in the region of \qtyrange{1}{100}{\milli\second}) and regimes where more solid-like properties dominate (typically longer than \qtyrange{10}{100}{\milli\second}). The crossover between solid and fluid regime depends on the cell type, and might reflect the biological function. Furthermore, on very long timescales that are in the order of the turnover of the cytoskeleton, the intracellular materials will become fluid again. The interested reader finds more details and further references to mathematically precise derivations in the \cref{tbox:shear_mod,tbox:microrheo,tbox:mech_models}.

\newpage
\begin{toolbox}[label=tbox:shear_mod]{Shear modulus}
    \setlength{\abovedisplayskip}{5pt}%
    \setlength{\belowdisplayskip}{5pt}%
    %
    Materials can exhibit different mechanical behaviors depending on how they respond to applied stress or strain. Elastic materials store energy when deformed and return to their original shape, while viscous materials flow under stress and dissipate energy as heat (\cref{fig:elastic viscous}). Viscoelastic materials combine both behaviors, showing features of solids and fluids at the same time, and in this case stress depends on both strain and strain rate. For example, corn starch or jelly behave elastic over short timescales (corresponding to high frequency in measurements where the deformation is oscillatory) and viscous over longer timescales (corresponding to low frequencies). Typically, this temporal sensitivity is reported as a frequency dependent property $(f)$.
    \begin{center}
        \begin{minipage}{\textwidth}
            \centering
            \captionsetup{type=figure}
            \includegraphics[width=\textwidth]{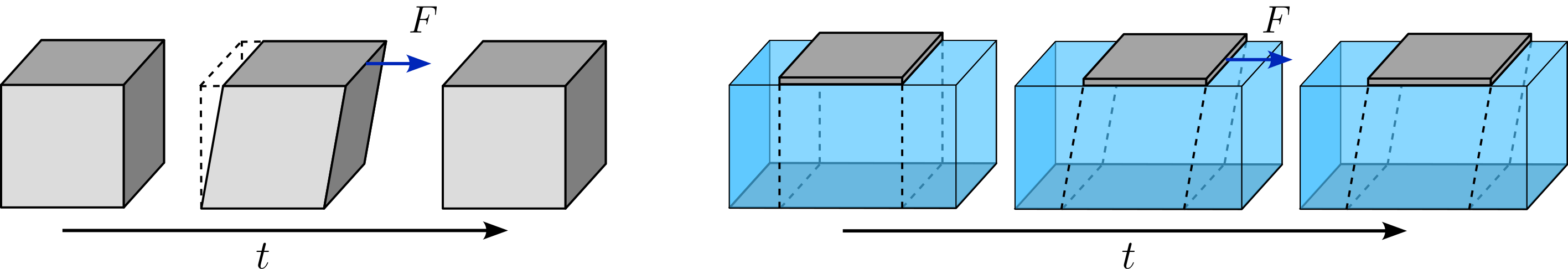} 
            \caption{\textbf{Elastic} materials return to their original shape when the stress is released (left sketch). In contrast, \textbf{viscous} materials flow under stress (e.g. applied by a metal plate). Depending on their viscosity, they may continue flowing (e.g. water) or stop at the position where the stress was released (e.g. ketchup, right sketch), but they never return to their original shape.} \label{fig:elastic viscous}
        \end{minipage}
    \end{center}

    To differentiate the elastic from the viscous properties, the elegant mathematical concept of complex numbers is used. The complex-valued frequency dependent shear modulus $G^{*} (f)$ characterizes the full viscoelastic material's response to a shear stress. This complex number has two components: The real part $G' (f)$, called the \textbf{storage modulus}, represents elastic behavior where deformation energy is stored; and the imaginary part $G'' (f)$, known as the \textbf{loss modulus}, reflecting viscous behavior where deformation energy is dissipated as heat. As mentioned, the viscoelastic properties vary across different time scales, making $G^{*}$ frequency-dependent. Thus, it is generally expressed as $G^{*} (f) = G^ {\prime } (f) + \mathrm{i} G^ {\prime \prime} (f) $, which highlights the necessity of probing the material over multiple time scales. Despite the mathematical complexity, this provides a straight forward and mathematically easy to handle description of the physics. More details can be found in the literature, \textit{e.g.} Chapter 4.5 in \cite{boal2012mechanics} and for full details Chapter 1--3 in \cite{lakes2009viscoelastic}.
    
    To determine $G^{*} (f)$, typically the complex mechanical response function of the material $\chi (t)$ is measured (\cref{tbox:microrheo}) and from there the complex shear modulus can be calculated using the geometry of the measurement device. The response function captures how a material reacts to forces $F(t)$ applied via a probe object, leading to deformations in the material. This response is reflected in the time dependent particle displacement $x(t)$, which in general depends on the entire history of forces applied to it (\cref{fig:response}). Thus, the mathematical relationship between force and displacement needs to include the full history of forces and is expressed as a convolution relation (\cref{eqn:response}). For a spherical probe particle of radius $R$, the relationship between $G^{*} (f)$ and Fourier transform of the response function $\tilde{\chi} (f)$ can be approximated by a Stokes term $6\pi R$ similar to the generalized Stokes-Einstein (GSE) relation (\cref{eqn:GSE}).  
    
    An alternative approach uses the system's memory function $\zeta (t)$, which is similar to $\chi (t)$ but relates force to the history of velocities (\cref{eqn:memory}). The Laplace transform of the memory function, $\hat{\zeta} (s)$, is connected to the viscoelastic spectrum $\hat{G}(s)$, and through analytical continuation (\cref{tbox:microrheo}), the shear modulus is obtained (\cref{eqn:GSE2}) \cite{squires2010fluid}.\\
    
    {
    \renewcommand{\arraystretch}{0.75}
    \begin{tabular}{>{\centering\arraybackslash}p{7cm}|
    >{\centering\arraybackslash}p{7cm}}
        Approach 1 & Approach 2 \\
        \begin{equation}\label{eqn:response}
            x(t) = \int_{-\infty}^{t} \chi(t-\tau) \, F(\tau) \,\mathrm{d}\tau        
        \end{equation}
        & 
        \begin{equation}\label{eqn:memory}
            F(t) = \int_{-\infty}^{t} \zeta(t-\tau) \, v(\tau) \,\mathrm{d}\tau
        \end{equation} \\
        \hline
        \begin{equation}\label{eqn:GSE}
            \text{Fourier-Transformed: } G^{*}(f) = \frac{1}{6 \pi R\, \tilde{\chi}(f)}
        \end{equation}
        & 
        \begin{equation}\label{eqn:GSE2}
            \text{Laplace-Transformed: } \hat{G}(s) = \frac{s \, \hat{\zeta}(s)}{6 \pi R}
        \end{equation}
    \end{tabular}%
    }
\end{toolbox}

\begin{toolbox}[label=tbox:microrheo]{Microrheology}
    \setlength{\abovedisplayskip}{5pt}%
    \setlength{\belowdisplayskip}{5pt}%
    \setlength\intextsep{\glueexpr\intextsep/2\relax}%
    Microrheology offers powerful methods to measure the complex shear modulus and to quantify mechanical properties at microscopic scales.
    This section highlights two main approaches relevant to determine the intracellular mechanics: active and passive microrheology.

    \vspace{0.2em}
    \textbf{Active intracellular microrheology} has its name due to the active external force that is applied to a probe particle inside a cell. Analyzing the particle's motion under these forces reveals the material's mechanical properties.
    For intracellular studies, oscillatory forces are applied to a particle in the cytoplasm using techniques like optical tweezers or magnetic traps (\cref{tbox:methods}) while recording force $F (t)$ and position $x (t)$. As introduced in (\cref{tbox:shear_mod}, \cref{eqn:response}), these quantities are related by a convolution, which drastically simplifies by taking the Fourier transform of \cref{eqn:response}. As a result, the complex response function can be readily solved and simplifies to $ \tilde{\chi} (f) = \tilde{x} (f) /  \tilde{F} (f)$. Here $\tilde {F} (f)$ and $\tilde{x} (f)$ are Fourier transforms of $F (t)$ and $x (t)$, respectively. Once the response function $\tilde{\chi} (f)$ is known, the shear modulus is readily accessible via \cref{eqn:GSE}. Moreover, the phase delay $\delta$ between $F (t)$ and $x (t)$ provides insights into the balance between viscosity and elasticity, given by $\tan(\delta) = G''(f)/G'(f)$. 
    \Cref{fig:microrheo_phase_delay} summarizes $\delta$ values for elastic, viscoelastic and viscous materials.
    
    \vspace{0.2em}
    \begin{minipage}{\textwidth}
        \centering
        \captionsetup{type=figure}
        \includegraphics[width=.7\textwidth]{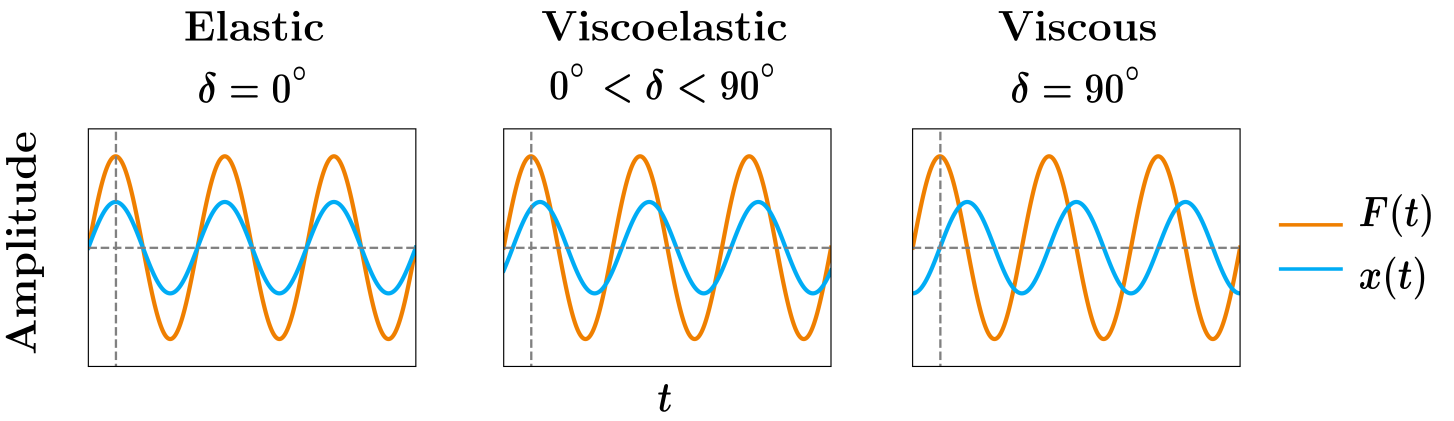}
        \caption{Phase delay between $F(t)$ and $x(t)$ for different types of materials.}
        \label{fig:microrheo_phase_delay}
    \end{minipage}
    \vspace{0.1em}
    
    \textbf{Passive microrheology} is simply passively observing a probe particle to infer mechanical properties. The cornerstone of this approach is the link between the memory function $\hat{\zeta} (s)$ (\cref{tbox:shear_mod}) and the mean squared displacement (MSD) \cite{squires2010fluid, mason1995optical, mason1997particle}. MSD depends on the average distance a particle moves from its starting point over time, and its relationship to $\hat{\zeta} (s)$ is expressed as
    \begin{equation} \label{eqn:shear_mod_MSD}
        \hat{\zeta}(s) = \frac{2\, k\_B T}{s^2} \frac{1}{\left< \Delta \hat{r}(s)^2 \right>},
    \end{equation}
    where $\left< \Delta \hat{r}(s)^2 \right>$ is the Laplace transform of the MSD. Therefore, by tracking a probe particle's trajectory $x(t)$ over time, $\mathrm{MSD}(t) = \left< |x(t) - x(0)|^2\right>$ is calculated, which allows determination of $\hat{\zeta} (s)$ using \cref{eqn:shear_mod_MSD}. From $\hat{\zeta} (s)$, the viscoelastic spectrum $\hat{G} (s)$ is determined using \cref{eqn:GSE2}. To obtain $G^* (f)$, either a fundamental physical assumption of causality (via Kramers Kronig relation) or a mechanical model (\cref{tbox:mech_models}) with real variable $s$, is applied to fit $\tilde{G} (s)$, providing an analytical expression. Finally, through analytical continuation (substituting $2 \pi \mathrm{i}\, f \rightarrow s$, Chapter 8 in \cite{ahlfors1979complex}), $G^* ( f )$ is obtained \cite{squires2010fluid, gittes1997microscopic, mackintosh1999microrheology}. 

    Passive microrheology, however, relies on the assumption that all motion is driven by thermal forces, which is generally not true in cells. Typically, applying passive microrheology approaches to living cell gives incorrect values for $G^*(f)$. However, by comparing passive and active microrheology, it is possible to determine the active forces with which cellular systems drive intracellular  \cite{hurst2021intracellular}.  
    
    It is important to add that the MSD can also be used to quantify a particle's diffusion. The MSD typically follows a power-law, where the prefactor is proportional to the diffusion coefficient $D$, indicating how fast a particle diffuses, and the exponent $\alpha$ characterizes the type of motion, as shown in the table below.

    \vspace{0.2em}
    \begin{minipage}{\textwidth}
        \centering
        \begin{tabular}{%
            @{}
            m{\linewidth/4-6\tabcolsep/4}|
            m{\linewidth/4-6\tabcolsep/4-\arrayrulewidth}|
            m{\linewidth/4-6\tabcolsep/4-\arrayrulewidth}|
            m{\linewidth/4-6\tabcolsep/4-\arrayrulewidth}
            @{}
        }
            $\mathrm{MSD}(t) = 6 D t^\alpha$&
            \includegraphics[width=\linewidth]{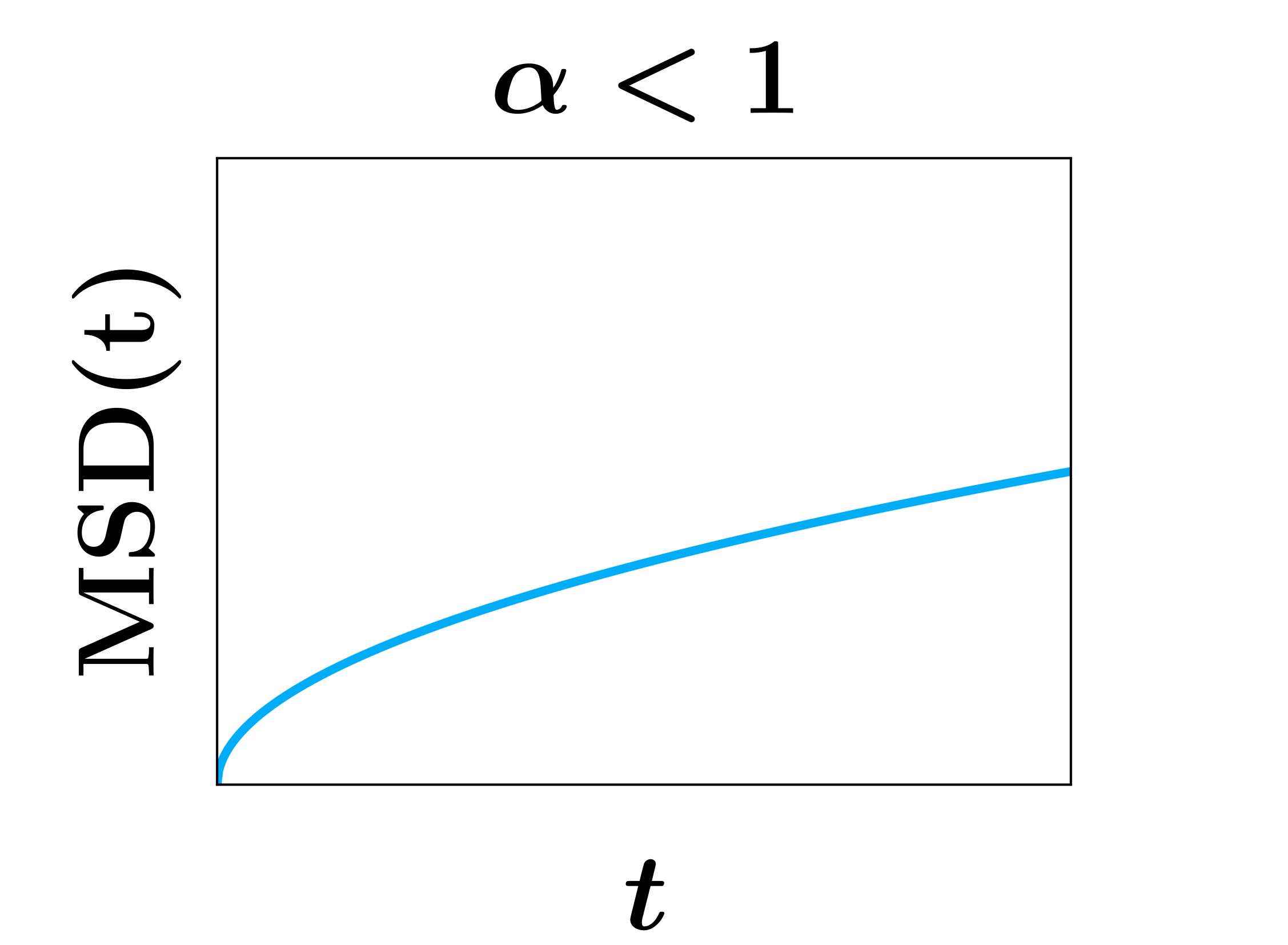} &
            \includegraphics[width=\linewidth]{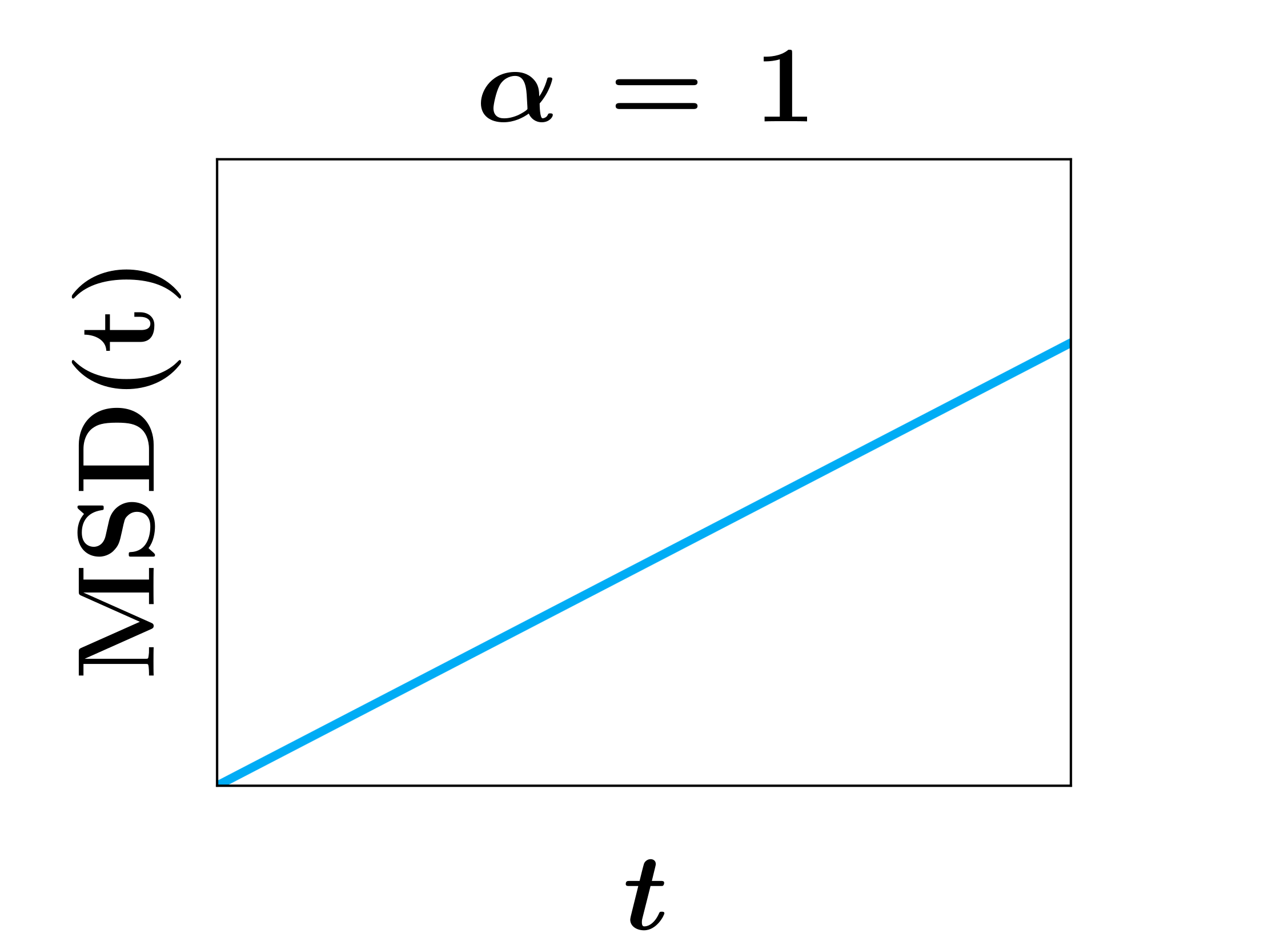} &
            \includegraphics[width=\linewidth]{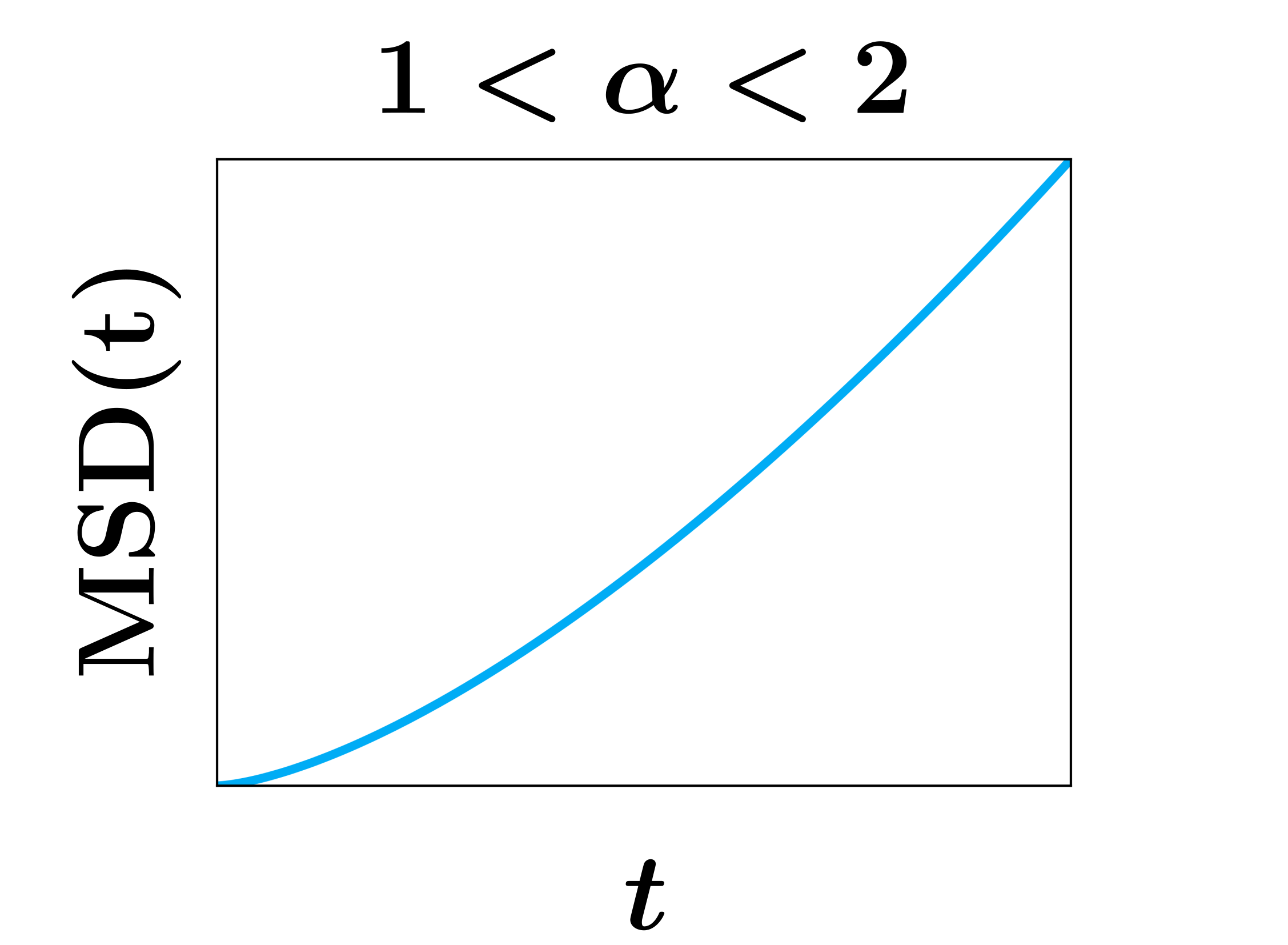}\\ \hline
            Type of motion & Sub-diffusive & Diffusive & Super-diffusive \\ \hline
            Example in \cref{fig:response} & Undriven cargo (3) & $\mathrm{Ca}^{2+}$ ions in cytosol & Motor-driven cargo (1)\\
        \end{tabular}%
    \end{minipage}%
\end{toolbox}

\begin{toolbox}[label=tbox:mech_models]{Mechanical Models}
    \setlength{\abovedisplayskip}{5pt}%
    \setlength{\belowdisplayskip}{5pt}%
    
    To interpret viscoelastic behavior, mechanical models are commonly employed to fit experimental data \cite{flugge1975viscoelasticity}. Among the various models available \cite{vos2024characterizing}, the Scott-Blair (SB) model stands out for its ability to capture the complex intracellular viscoelastic properties of cells over a broad frequency range \cite{blair1944study,bonfanti2020fractional, nguyen2021multi}. Mathematically, the complex shear modulus in SB model is expressed as
    \begin{align}
        G^*(f) &= A \left( 2 \pi \mathrm{i} \, \frac{f}{f_0} \right)^{\alpha},
    \end{align}
    where A determines its strength, and $\alpha$ specifies its position on the spectrum between purely elastic $\left( \alpha = 0 \right)$ and purely viscous $\left( \alpha = 1 \right)$ behavior. To ensure dimensional consistency, the frequency is normalized by $f_0$, which is conventionally set to \qty{1}{\hertz}. However, experimental data may deviate from this model at very low frequencies  $\left( f < \qty{0.1}{\hertz}\right)$ \cite{chaubet2020dynamic}. Furthermore, the physical interpretation of the fitting parameters $A$ and $\alpha$ remains an area of ongoing investigation.
    
    Since materials like the cytoplasm often exhibit distinct mechanical responses at low and high frequencies (\cref{fig:shear_mod}), employing two SB elements in parallel allows for a more accurate representation of intracellular mechanics. This is called fractional Kelvin-Voigt model and is mathematically expressed as
    \begin{align}
        G^*(f) &= A \left( 2 \pi \mathrm{i} \, \frac{f}{f_0} \right)^{\alpha} + B \left( 2 \pi \mathrm{i} \, \frac{f}{f_0} \right)^{\beta}.
    \end{align}
    \vspace{0.3em}
    \begin{minipage}{\textwidth}
        \centering
        \captionsetup{type=figure}
        \includegraphics[width=.35\textwidth]{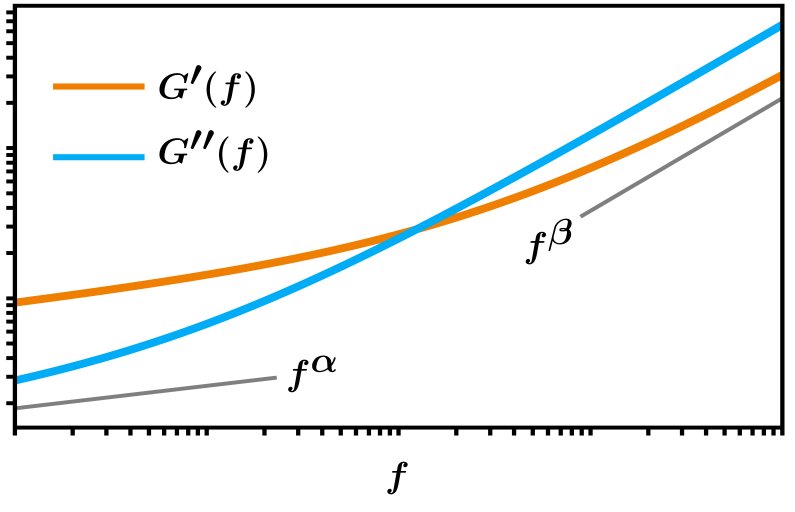}
        \caption{$G^\prime$ and $G^{\prime \prime}$ show power law behavior, but with different exponents in low and high frequencies.}
        \label{fig:shear_mod}
    \end{minipage}
\end{toolbox}%

\begin{center}
    \begin{minipage}{\textwidth}
        \centering
        \captionsetup{type=figure}
        \includegraphics[width=0.5\textwidth]{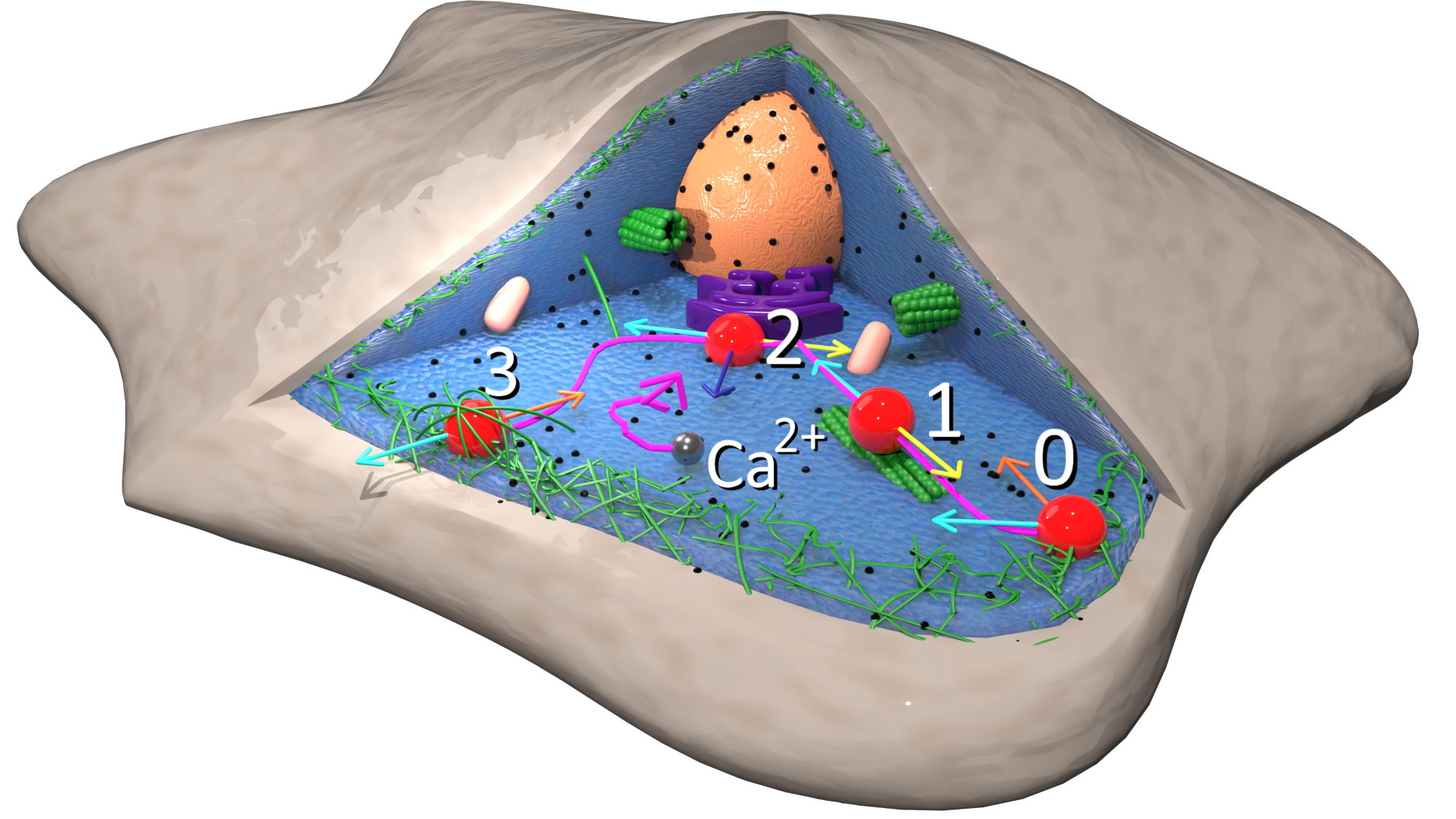}
        \caption{In general, the position of a particle (red sphere) does not only depend on the forces $F (t)$ acting on it in right now, but on the force history. All forces acting along its path from position 0 to position 3 contribute to its state at position 3.
        The environment's response to applied forces is characterized by the \textbf{response function} $\chi(\Delta t)$, which reflects the material's mechanical properties. Intuitively, $\chi (\Delta t)$ describes the particle motion resulting from a force that has been applied $\Delta t$ seconds before. In this figure, forces arising from motors, actin flow, drag, and organelle pushes are represented by $F_\mathrm{M}$ (cyan arrows), $F_\mathrm{A}$ (orange arrows), $F_\mathrm{D}$ (yellow arrows) and $F_\mathrm{O}$ (dark blue arrow), respectively.
        \\ 
        Moreover, the motion type at each position is determined by the mechanics there. It can be \textbf{sub-diffusive} as a result of passive or active confinement within the cytoskeleton (position 3), \textbf{diffusive} (e.g. $\mathrm{Ca}^{2+}$ freely diffusing in the cytosol), and \textbf{super-diffusive} as a result of active transport (e.g. position 1).
        }
        \label{fig:response}
    \end{minipage}
\end{center}
\begin{toolbox}[label=tbox:methods]{Measurement methods to determine cell mechanics}
\pdfoutput=1

    Cells are very small and soft objects, hence measuring mechanical properties inside cells requires sophisticated instruments \cite{hao2020mechanical}. Tools like atomic force microscopy (AFM) \cite{krieg2019atomic, rother2014atomic} and micropipette aspiration \cite{hochmuth2000micropipette} give the mechanical properties of the cell's outer surface and the underlying cortex.
    Magnetic \cite{bausch1999measurement, wang2019intracellular} or optical tweezers \cite{brau2017passive, volpe2023roadmap, catala-castro2022exploring}, as well as particle tracking microrheology (PTM) \cite{ruthardt2011single} enable researchers to explore the distinct mechanical properties and particle mobilities within the cell, which are in general drastically different to the cortex mechanics.

    \vspace{0.5em}
    \begin{minipage}{\textwidth}
        \centering
        \captionsetup{type=figure}
        \includegraphics[width = 0.69\textwidth]{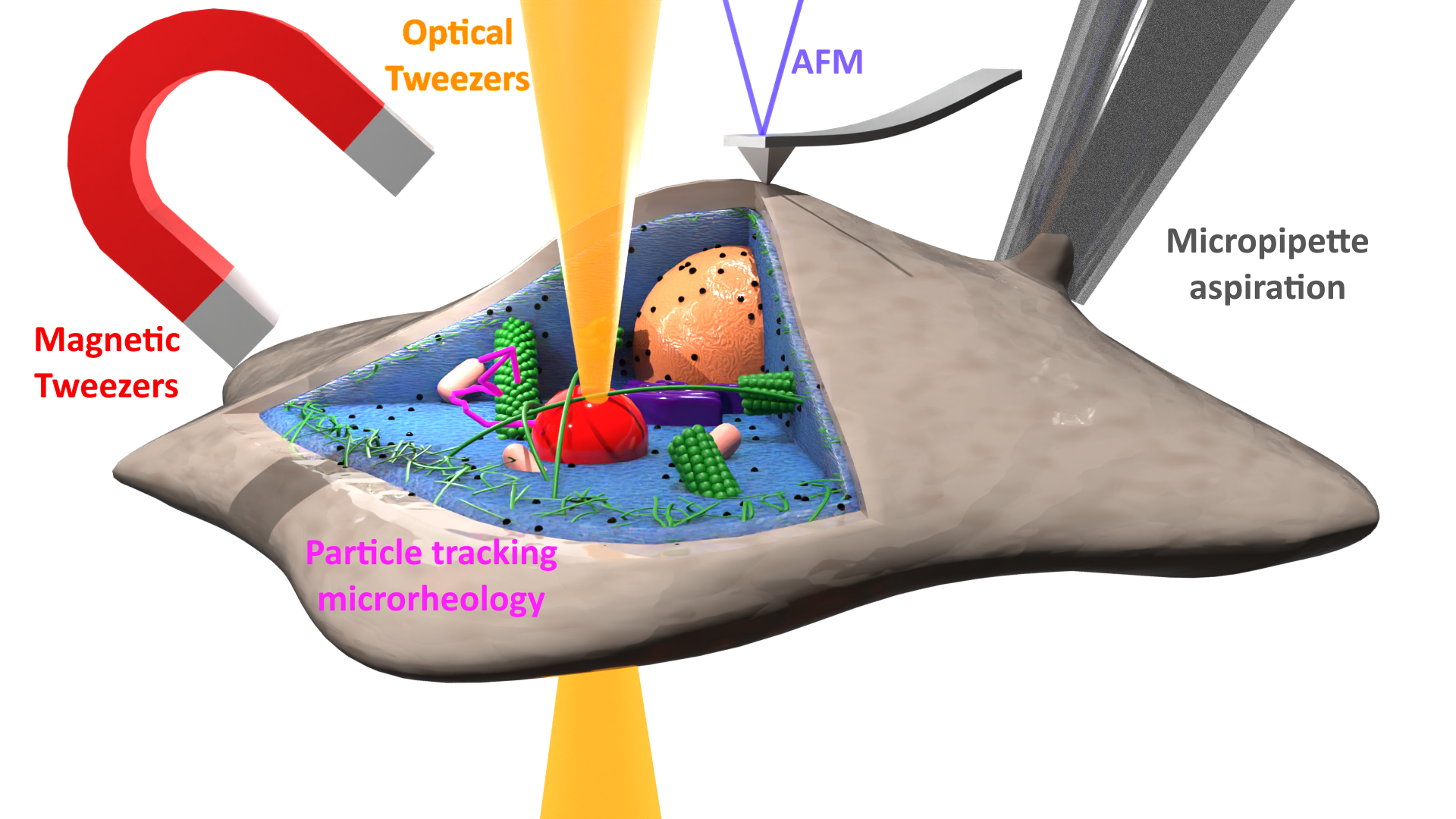}
        \caption{%
            Illustration showing the measurement techniques used to obtain cell mechanics.  
            Atomic force microscopy (AFM) and micropipette aspiration are techniques that usually probes the whole-cell mechanics and activity.
            Methods like optical tweezers, magnetic tweezers and particle tracking microrheology allows an insight on the intracellular mechanics.
        }
        \label{fig:method_big}
    \end{minipage}
    
    \begin{tcbraster}[raster columns=2, raster equal height]
        \setlength{\tabcolsep}{2pt}
        \begin{innertoolbox}[]{External measurements}%
            \begin{minipage}{0.25\textwidth}
                \includegraphics[width=\linewidth]{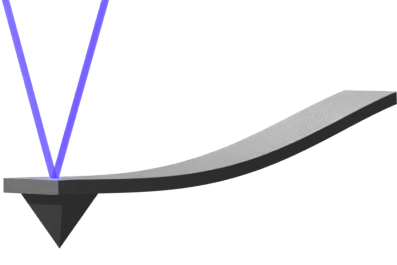}
            \end{minipage}
            \hfill
            \begin{minipage}{0.7\textwidth}
                Since its invention in 1986 \cite{binning1986atomic}, \textbf{AFM} has become one of the most implemented techniques used in cell mechanics to measure cells' elasticity and rheological properties \cite{moeendarbary2014cell}.
                A small tip is attached to a flexible cantilever that will probe the cell's surface.
                This interaction results in the bending of the lever, thus deflecting the laser light from the back of the lever onto the photodiode.
                Force-indentation curves are obtained from such experiments, out of which can be extracted the cell's stiffness (Young's modulus) via fitting contact models (Hertz, Sneddon, Oliver and Pharr) \cite{dufrene2017imaging, kontomaris2020hertz}.
            \end{minipage}
            
            \vspace{0.5em}
            
            \begin{minipage}{0.25\textwidth}
                \includegraphics[width=\linewidth]{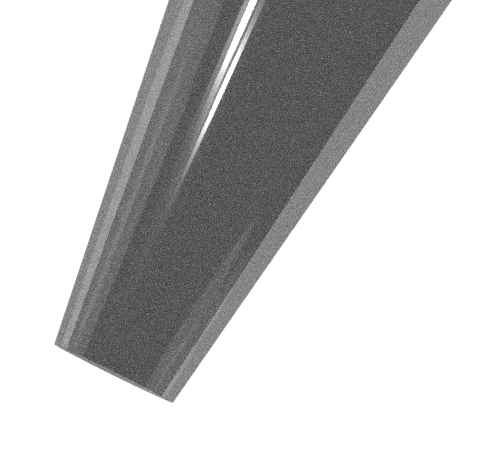}
            \end{minipage}
            \hfill
            \begin{minipage}{0.7\textwidth}
                \textbf{Micropipette aspiration} is another technique used to measure cell mechanics.
                The cell membrane and the cortex are sucked into a glass pipette at a known applied pressure.
                From the length of the cell projection and the pressure, a linear regression can be obtained where the slope yields the Young's modulus \cite{hochmuth2000micropipette, theret1988application, khani2014evaluation}.
            \end{minipage}
        \end{innertoolbox}
        \begin{innertoolbox}[]{Internal measurements}
            \begin{minipage}{0.7\textwidth}
                The \textbf{optical tweezers} consist of a focused laser beam which, thanks to the applied gradient force, keeps a particle in the centre of the beam. It can be used to apply well defined minuscule forces on particle inside cells and analyse thereby their mechanical properties \cite{brau2017passive, volpe2023roadmap, catala-castro2022exploring, ashkin1986observation}.
            \end{minipage}
            \hfill
            \begin{minipage}{0.25\textwidth}
                \includegraphics[width=0.43\linewidth]{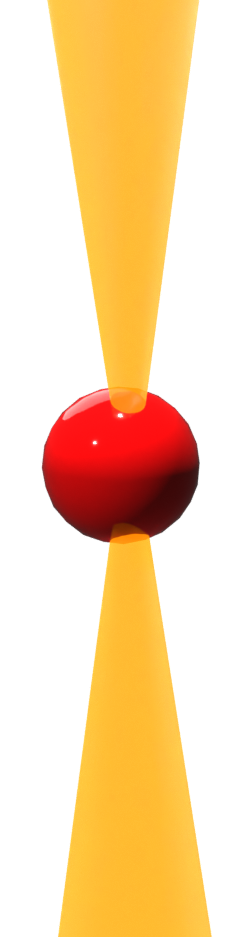}
            \end{minipage}%
            
            \vspace{0.5em}
            
            \begin{minipage}{0.7\textwidth}
                \textbf{Magnetic tweezers} are a tool for manipulating the inside of living cells or measuring the intracellular mechanical properties.
                Similar to optical tweezers, particles inside the cells are moved by the gradient of a magnetic field. The small forces leads to well defined particle motion which is used to determine the intracellular mechanics \cite{bausch1999measurement, wang2019intracellular}.
            \end{minipage}
            \hfill
            \begin{minipage}{0.25\textwidth}
                \includegraphics[width=0.9\linewidth]{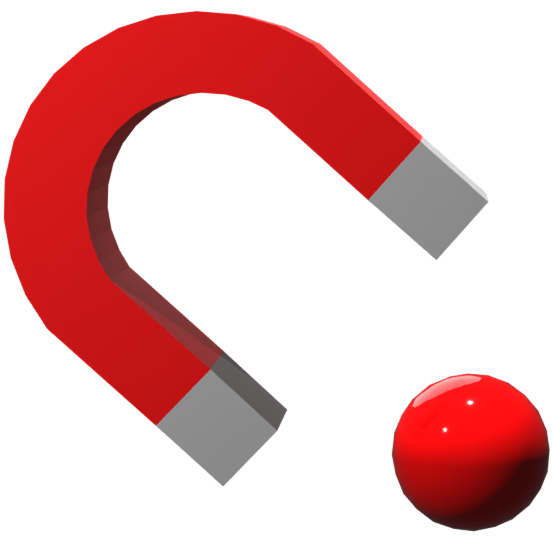}
            \end{minipage}%
            
            \vspace{0.5em}
            
            \begin{minipage}{0.7\textwidth}
                In \textbf{particle tracking microrheology (PTM)}, particles are tracked within cells with high spatiotemporal resolution.
                Although passive movement cannot be distinguished from active motion, it is sometimes assumed that no active forces are applied when determining viscoelastic properties \cite{ruthardt2011single, wirtz2009particle}. Typically, this assumption results in a drastic underestimation of intracellular stiffness and should not be used.
            \end{minipage}
            \hfill
            \begin{minipage}{0.25\textwidth}
                \includegraphics[width=0.9\linewidth]{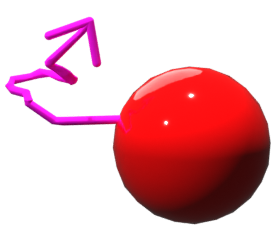}
            \end{minipage}
        \end{innertoolbox}
    \end{tcbraster}
\end{toolbox}
\section{Intracellular mechanics is the result of intracellular content}\label{sec:constituents}
\pdfoutput=1

To appreciate the importance of intracellular mechanics, we will first review the molecules relevant for generating and controlling intracellular mechanical properties. It is important to realize that the cytoplasm is mostly made up of water (\qty{70}{\percent} of mass) \cite{neurohr2020relevance} and the remaining material is referred to as \textit{dry mass}. Of course, water is very relevant for proper function, but since cells are mechanically very different from water, this dry mass must play an important role for the mechanical properties of the cell.
Its most abundant component are proteins (about \qtyrange{50}{60}{\percent} of dry mass \cite{alberts2002chemical,fischer2004average}) whereas carbohydrates, lipids and RNA take second place depending on the cell line \cite{neurohr2020relevance}.
A main focus of cell biology is to assign a role and function to the proteins inside a cell.
Only a subclass of these is known to be important for the cellular mechanics.
Here, we focus on the proteins that have a known direct or indirect effect on intracellular mechanics: The cytoskeleton and their associated proteins.
Motor proteins---while important---are not the focus of this section, but the activity they supply to the cell is discussed in \cref{sec:activity}.
Other relevant topics that cannot be covered here but are discussed in the literature, include mechano\-sensing and mechano\-transduction (how is mechanical stress perceived by the cell and converted into a biochemical signal) \cite{staunton2019high,mathieu2019intracellular}, modifications of transcription due to mechanical perturbation \cite{tajik2016transcription}, size and speed dependence of mechanical properties \cite{hu2017size,najafi2023size} as well as the higher level mechanics of tissues \cite{trepat2020measuring}.

\subsection{Cytoplasmic density and crowding}
\label{sub:density_crowding}

Often underestimated, the volume regulation of cells is a highly relevant factor for mechanical properties, as it affects the protein density. Such volume changes are typically driven by the physical effect of osmotic pressure that results in water in- or outflow through the membrane and its pores.
As mentioned in the introduction, the high density of proteins results in important molecular crowding. Due to the extent of crowding, it is believed that the cytoplasm is close to a glass or jamming transition where local compartments are mechanically arrested and active energy plays an important role to fluidize the cytoplasm \cite{mizuno2017universal} (or stiffen it---see the discussion in \cref{sec:activity}).
Crowding is of increasing interest due to its relevance for transport \cite{nettesheim2020macromolecular}, physiology \cite{mourao2014connecting}, and microtubule polymerization \cite{molines2022physical}.
Consistent with its relevance, an increasing number of studies suggests that the density of the cytoplasm is tightly regulated with cell volume playing a crucial role \cite{neurohr2020relevance, biswas2024exploring, chen2024viscosity}.
Cell volume is reported to be related to mechanical properties \cite{guo2017cell} with larger cells being softer on average.
This effect of density on viscosity and mechanics is quite independent on the actual proteins themselves, but remains a physical crowding effect that is currently under heavy investigation. 

Albeit important, of course it is not just the overall protein density that controls intracellular mechanics.
In reality we know very well that intracellular mechanics is largely controlled by a special class of proteins that constitute the cytoskeleton.

\subsection{Cytoskeletal filaments}\label{sub:cytoskeletal-filaments}

The cytoskeleton is mainly composed of three types of filaments: Actin filaments, microtubules and the class of intermediate filaments.

\subsubsection*{Actin filaments}
Actin structures such as lamellipodia, filopodia, contractile stress fibres and the cell cortex combine to enable cell migration, sensing and provide cell shape \cite{blanchoin2014actin}. An important feature of the actin cytoskeleton is its ability to create an impressive broad variety of different structures.
These depend not only on actin monomers concentration, nucleators and crosslinkers but even more on regulatory proteins that control and adjust the viscoelastic properties of the resulting networks (tuning nucleation, branching, capping and depolymerization) \cite{blanchoin2014actin,lappalainen2022biochemical}.
Besides these biochemical adjustments, the local mechanical environment plays an additional, often overlooked role in the actin structures. An impressive example of this interaction between local mechanics and structure is the intrinsic self-regulatory feedback that allows cell to adjust the polymerization based pushing forces during migration. Imposing mechanical stress on the growing region of an actin network changes the polymerization dynamics that are mediated by actin binding proteins, typically slowing down growth while increasing branching and thereby network density \cite{lappalainen2022biochemical}. This in turn increases the stability and the pushing power of the network, which allows to counteract the mechanical stress that led to these changes \cite{bieling2016force}. \textit{In vitro} experiments have demonstrated the impressive ability of actin networks to vary stiffness from values as low as \qty{0.1}{\pascal} \cite{gardel2003microrheology} up to \qtyrange{1}{10}{\kilo\pascal} \cite{bieling2016force, chaudhuri2007reversible, marcy2004forces, pujol2012impact,ruckerl_adaptive_2017}.
Besides these biomechanical feedback regulations, actin networks are also known to be a highly nonlinear material, which inherently stiffens when pulled beyond one critical point and reversibly softens afterwards \cite{chaudhuri2007reversible, fletcher2010cell}.
Consequently, the actin cytoskeleton is found to be critical for the overall cell mechanics that is largely fulfilling this biomechanical role at the cell cortex. However, it comes to a big surprise, that actin is almost irrelevant for the intracellular mechanics of many cells when measured using optical tweezers. Several studies have demonstrated that the viscoelastic properties inside a cell, hence away from the cell cortex, are hardly changing upon disruption of the actin cytoskeleton \cite{ahmed2018active, hurst2021intracellular, muenker2023intracellular, ebata2023activity}. In turn, in starfish oocyte experiments testing intracellular mechanics using magnetic tweezers on nuclear sized probes, actin depolymerization softened the cytoplasm \cite{najafi2023size}. 
Intriguingly, actin seems to be a key player for cellular properties on the whole cell scale via the cell cortex, but is not found to be similarly important for the intracellular mechanics. 

\subsubsection*{Microtubules}
Besides actin, microtubules are known to be a further key cytoskeletal filament structure inside cells. Microtubules play important roles in trafficking, signalling, morphogenesis \cite{matis2020mechanical} and mitosis. While their mechanical properties have been extensively studied \textit{in vitro} \cite{gudimchuk2023measuring, kerssemakers2006assembly,}, they have been found not very relevant for the global cell stiffness, which is dominated by the actin cortex. 
Similar to actin, only few studies have investigated the effect of microtubules on intracellular mechanics. Surely, in the cellular region of the spindle either in oocytes or during cell division, microtubules are considered a stiff local structure. Furthermore, by optical tweezers based active microrheology, microtubules have been identified as the dominating cytoskeletal structure inside HeLa and MDCK cells, providing both active transport forces and mechanical rigidity \cite{hurst2021intracellular, muenker2023intracellular}. Interestingly, in starfish oocytes, inhibiting microtubule polymerization let to a stiffening of the cytoplasm, suggesting either a softening role, or that microtubule depolymerization triggers signalling cascades that eventually stiffen the cells. This could be explained by an increase in actin filaments via the release of microtubule sequestered rhoGTPases to compensate the loss in mechanical integrity provided by the microtubules \cite{chang2008gef-h1, krendel2002nucleotide}, showing the interconnected properties of the different cytoskeletal components.
As later established, this interplay of actin and microtubules plays a role in numerous cellular functions \cite{dogterom2019actin}.

\subsubsection*{Intermediate filaments}
While actin and microtubules have been thoroughly studied in the past three decades, intermediate filaments are much less understood since the technical obstacles to research have only been overcome in the past 5-10 years.
Single intermediate filament \textit{in vitro} studies establish this family of filaments as highly resilient to mechanical perturbations and as a key player in stress dissipation \cite{buehler2013mechanical, roelleke2023unique}.
This role is also found studying wild type, vimentin knockout and vimentin ghost cells (keeping only the vimentin network of a former wild type cell) where stretchability and resilience of the cells to mechanical strain were mainly provided by the vimentin network \cite{hu2019high}. 
Furthermore, the hyperelastic characteristic of the vimentin network alone does not dissipate energy at moderate to large deformations (in contrast to single filaments that mostly dissipate \cite{roelleke2023unique}).
However, in conjunction with actin filaments and microtubules the dissipation of deformation energy is greatly increased, allowing the cell to survive large deformations \cite{hu2019high}.
This interaction between intermediate filaments and the other cytoskeletal components, actin and microtubules, seems to be very relevant for a complete picture, as recently confirmed both \textit{in vitro}  \cite{shen2021effects, pradeau-phelut2024cytoskeletal}, but also in cells where actin and intermediate filaments have been found to create interconnected networks \cite{wu2022vimentin}.
Notably, due to their non-polar structure, no motor protein can persistently move along intermediate filaments \cite{fletcher2010cell}.
Intermediate filaments are therefore expected to contribute less to the active processes in cells if neglecting the interplay with other filaments and their associated proteins. Still, their well controlled localization suggests a relevant role for intracellular mechanics that has yet to be precisely defined.

\subsection{Associated proteins and signalling}\label{sub:associated-proteins}

Associated proteins interact with the polymerization characteristics (providing nucleation sites, modifying binding/unbinding rates), cross-link filaments, link filaments to other (\textit{e.g.} membranous) structures and provide intracellular signalling. Best studied are actin associated proteins, although microtubule associated proteins (MAPs) are known to be of immense importance for proper cell function. In turn, intermediate filaments continue to be a blind spot also in terms of knowledge regarding their associated proteins.
Besides proteins that were established to directly regulate \textit{e.g.} the filament polymerization, different classes of higher order signalling proteins (proteins regulating other regulators) were uncovered \cite{hohmann2019cytoskeleton}.
The mechanical properties of the cytoskeleton therefore present themselves as being dependent not only on the composition of the three main filament types, but a full mechanical picture of the cytoskeleton needs to also cover its associated proteins and their downstream targets \cite{moujaber2020cytoskeleton}.  
This results in a highly complex and dynamic architecture that adapts to external and internal cues (signalling).
\textit{In vitro} studies so far are only starting to incorporate associated proteins to understand their mechanical role in a well-controlled system \cite{landino2021rho, thery2024reconstituting, litschel2021reconstitution}.

Turning to concrete examples, 
Rho GTPases are the best-studied family of proteins for actin that influences cell tension and cytoskeletal organization \cite{andersen2023cell}.
In turn, actin binding proteins (ABPs) provide a versatile toolbox for controlling cellular mechanical properties \cite{vakhrusheva2022role} and have been extensively studied in reconstituted systems. 

Microtubule associated proteins (MAPs) can stabilize, facilitate rescue, persist the growth of the filament and reduce the frequency of catastrophes \cite{hohmann2019cytoskeleton}.
Similar to actin, the class of Rho GTPases represents the most notable family of regulating proteins for microtubules and MAPs  \cite{hohmann2019cytoskeleton}.

In contrast, knowledge about regulation and associated proteins for intermediate filaments remain rather sparse. However, as intermediate filaments and their mechanical role is an emerging research focus, one might expect their most relevant associated proteins to be discovered in the near future.

For a more detailed account of the interacting cytoskeleton including the interactions of its components and a list of associated proteins the reader is kindly referred to the up to date review \cite{hohmann2019cytoskeleton}.

\subsection{Intracellular Mechanics Under Cytoskeletal Control}

In summary, although a lot is known about the impact of the cytoskeleton on the overall cell mechanics, we have only very superficial insights how and to which extend the cytoskeleton controls the intracellular mechanics. In turn, stiffness and moreover viscosity of the intracellular space has drastic effects on transport properties, not only for organelles and vesicles, but also for diffusion which can limit biochemical reaction rates.
Conversely, signalling and associated proteins are expected to control the local mechanical properties of the cytoplasm, leading to reciprocal interactions that remain poorly explored.
Indeed, the importance of the proteins associated with the cytoskeleton is not yet well reflected in studies on intracellular mechanical properties.
While reconstituted \textit{in vitro} studies are starting to gain complexity and investigate the interactions of two types of networks, the combination with biochemical signalling remains a difficult task. A better understanding here could be achieved combining bottom-up and top-down approaches. 
The bottom-up strategy followed in reconstituted systems can increase the complexity to converge to the more detailed situation of cultured cells, while remaining illustrative of mechanics.
In turn, top-down strategies reduce the inherent complexity of cellular systems by the means of knockout cell lines.

Thanks to the protein-based studies of the cytoskeleton, it is evident that intracellular mechanics depends on the spatial (see \cref{sec:spatial_variations}), but also temporal distribution of the cytoskeleton (see \cref{sec:temporal_changes}), and other intracellular structures, such as membranes and organelles. Furthermore, the influence of active forces that are continuously rearranging the intracellular space must be considered (see \cref{sec:activity}). In light of this complexity, it remains a mystery how cellular systems manage to organize and stabilize their intracellular structures, mechanical properties and active non-equilibrium forces.  
\section{Spatial variability of intracellular mechanics}
\label{sec:spatial_variations}
\pdfoutput=1

To appreciate the difficulties living cells have to overcome to achieve their stable internal organization, it is helpful to recall that in equilibrium systems, small structures have a strong tendency to disperse simply by diffusion \cite{joshi2024emergent}. This is the reason why dead cells disintegrate over time. To overcome these ever dispersing entropy based forces \cite{mukhopadhyay2008lipid}, living system have developed an impressively effective system to keep global spatial organization not only by transport but also by spatial heterogeneity of intracellular active viscoelastic properties, which we only begin to understand. 

The cytoskeleton, in particular actin and microtubules, plays a central role in the positioning of organelles in cells. This is accomplished through three primary mechanisms: (1) structural heterogeneity in the cytoskeletal network, which provides stabilization and reorganization \cite{xie2022contribution, minc2011influence}, (2) transport and anchoring of organelles by motor proteins like dynein and kinesin, ensuring precise localization \cite{yadav2011golgi, akhmanova2022mechanisms}, and (3) active diffusion that not only enhances the mobility of small organelles, but also positions large organelles, such as the nucleus, in their correct locations \cite{colin2020active, almonacid2015active}. 

This dynamic yet non-random positioning of organelles \cite{schauer2010probabilistic, joshi2024emergent}, helps the cells to overcome diffusion limitations by positioning organelles closer to the sites where their biological functions are required. For instance, in migrating cells the Golgi apparatus is positioned towards the front of the cell to ensure efficient delivery of newly synthesized proteins to the cell cortex, while endosomal compartments align with the microtubule-organizing center (MTOC) for protein recycling \cite{fourriere2019rab6, olenick2019dynein}. On the molecular organization level, liquid-liquid phase transitions have been shown to use fundamental physical effects to provide further local organization, and the relation between these droplets and the local viscoelastic properties is under heavy investigation \cite{berry2018physical, vidal2020theory, kothari2023crucial}. Each of the above listed spatial organization mechanisms are affected by mechanical properties, as will be discussed later in this section. 

\subsection{Coordination between cytoskeletal heterogeneity and mechanical properties}

Cellular mechanics vary across different regions of the cytoplasm \cite{mclaughlin2020spatial}. 
For example, tracking individual genetically encoded multimeric nanoparticles (GEMs) reveals a 400-fold variation in their apparent diffusion coefficient \cite{garner2023vast}. This variation in general correlates with the heterogeneity in cytoskeleton organization and organelle positioning. For instance, when cells are plated on micropatterns that regulate their shape, a distinct pattern in the stiffness, as quantified by the shear modulus (see \cref{tbox:shear_mod}), emerges. Radial and front-rear polarity become evident, with the shear modulus decreasing from the cell center toward the periphery. This decrease in stiffness aligns with a reduction in the density of vimentin filaments. Additionally, the polarization from rear to front reflects the underlying organization of the cytoskeleton and key organelles, including the Golgi apparatus and the endoplasmic reticulum \cite{mandal2016mapping}.

\medskip

Cytoskeletal heterogeneity and organelle positioning are not merely statically coordinated; changes in mechanical properties can also influence the repositioning of organelles. An impressive experimental evidence for this is provided by applying forces to a cell using an AFM tip (see \cref{tbox:methods}), which then triggers significant spatial rearrangements of mitochondria, with local forces displacing organelles even at the cell's edge, far from the initial point of contact \cite{silberberg2008mitochondrial}. This effect is largely driven by the mechanical properties of cytoskeletal filaments, meaning that heterogeneity in these properties determine how far external forces can propagate through the cytoplasm, thereby affecting organelle distribution. In addition, by inducing anisotropy to cell morphology, it has been shown that the cytoskeletal filaments align in the elongated direction and this in turn makes the cytoplasm stiffer along the same direction \cite{gupta2019anisotropic}. It is interesting that this anisotropy in stiffness is also reflected in the motion of endocytosed microbeads, suggesting that cells might reposition organelles by adjusting their mechanical properties \cite{gupta2019anisotropic}.  Important to note is that althought the cytoskeleton is the primary contributor to intracellular mechanics, the role of membranous structures and their interaction with the cytoskeleton should not be overlooked. For instance, studies have shown that the Golgi apparatus is stiffer than the surrounding cytoplasm, with the actin cytoskeleton playing a critical role in reinforcing the rigidity of Golgi membranes and the peri-Golgi region \cite{ravichandran2020golgi}. However, this aspect remains underexplored, highlighting the need for further investigation.

\subsection{Mechanical stabilization by cytoskeletal heterogeneity}

The coordination between cell mechanics and organelle positioning suggests that viscoelastic properties might be utilized by cells to stabilize the position of the organelles. For instance, large objects, such as the cell nucleus, can experience anisotropic and position-dependent viscoelasticity within the cytoplasm due to hydrodynamic interactions with the cell surface. This effect may restrict the movement of the nucleus toward the cell boundary, helping to maintain its central positioning. \cite{najafi2023size}. Moreover, it is well known that lysosomes are unevenly distributed in the cytoplasm and their spatial distribution and positioning relative to one another remain stable. This stability is achieved by the constraining effects of the actin network on diffusion and the balance between long-range centrifugal and centripetal transport facilitated by microtubule related motors \cite{ba2018whole}. This illustrates how cells can integrate multiple mechanisms---such as spatial variation in mechanics and active transport---to organize their intracellular environment.

\subsection{Effect of mechanical properties on intracellular transport}

Mechanical heterogeneity not only plays a role in stabilizing and repositioning organelles, but also significantly influences the transport of organelles and cargo. In crowded regions of the cytoplasm, where stiffness increases, cargo may either recruit additional motor proteins to overcome the viscous drag of the cytoplasm \cite{wieland2022sizing} or experience reduced movement depending on their size \cite{luby1986probing}. The influence of mechanical properties on transport is reflected in changes to the diffusion coefficient and the type of motion exhibited by organelles and cargo (see \cref{tbox:microrheo}). A perfect example for this is mitochondria transport, that exhibits a non-uniform distribution of both diffusive and super-diffusive motion with the diffusion coefficient varying by more than \qty{50}{\percent} \cite{bomzon2006mitochondrial}. A stunning example of mechanical effects on mobility is lysosome transport. These organelles also show three distinct populations, each exhibiting different modes of movement: sub-diffusive, super-diffusive, and diffusive, however their relative distance remains constant over time. Although this seems to contradict each other, it can be easily explained by local heterogeneity of mechanical properties  \cite{ba2018whole}. Also, lipid granules seem to reveal local variation in intracellular structure as their motion depends on their localization within the cytoplasm. Granules located at the periphery, where actin is abundant, move less than those in the central region of the cell \cite{leijnse2012diffusion}. Similarly, phagocytosed microspheres exhibit super-diffusive motion near the nucleus, attributed to interactions with microtubule-associated motor proteins \cite{caspi2000enhanced}. These findings suggest that organelle transport can be governed by the spatial variation in viscoelastic properties, such as elasticity and viscosity, which in turn determine the diffusion coefficient and motion characteristics.

\subsection{Active diffusion gradients maintain intracellular organization}

Although active transport and cytoskeletal heterogeneity are essential for intracellular organization, additional mechanisms further enrich the cellular toolkit. Increasing evidence highlights alternative strategies for organelle positioning that do not rely on anchoring organelles to the cytoskeleton \cite{huelsmann2013filopodia, moore2021actin, illukkumbura2020patterning}, but rather depend on mechanical properties. One such mechanism is active diffusion, where random motor-driven motions lead to purely diffusive behavior over long timescales \cite{brangwynne2009intracellular}. This process can generate propulsive forces substantial enough to reposition large organelles like the nucleus, typically steered then by gradients of this active diffusion. For example, in mouse oocytes, nuclear positioning is driven by the active diffusion of actin-coated vesicles, facilitated by a reduction in both the storage and loss moduli (see \cref{tbox:shear_mod}), effectively fluidizing the cytoplasm \cite{almonacid2015active}. This process is able to position any object of similar size, showing that it is a simple physical effect, not relying on additional signalling, rendering it highly robust \cite{colin2020active}. While mouse oocytes do not exhibit spatial variation in their intracellular mechanical properties, this mechanism of active diffusion gradients might be also relevant in somatic cells. We propose that active diffusion, combined with mechanical heterogeneity, could provide cells with an even more versatile and robust means of intracellular organization, making it a promising area for future research.

\section{Temporal changes in intracellular mechanics }
\label{sec:temporal_changes}
\pdfoutput=1

Although spatial variations of intracellular mechanics are thought to be important, emerging evidence suggest that temporal changes of intracellular active mechanics can provide insight in general cell biological questions. The link between (intra-)cellular mechanics and biological function has become increasingly evident in recent years, with studies demonstrating how physical parameters like cell volume and stiffness can profoundly influence key processes such as differentiation \cite{guo2017cell}. Moreover, a rapid increase in volume during mitosis leads to a decrease in protein density of the cytoplasm, which explains the decrease in intracellular stiffness\cite{hurst2021intracellular}, thereby facilitating chromosome transport during division \cite{son2015resonant, zlotek2015optical}. All of these findings suggests a pivotal role of (intra-)cellular mechanics during temporally dynamic biological processes, including embryogenesis, cell division, and cellular differentiation. Indeed, reviewing the temporal changes of intracellular mechanics in relation with cellular function reveals an intricate interplay between mechanics and function in shaping biological systems.

\subsection{Temporal evolution of intracellular mechanics during embryogenesis}

Mechanics play a crucial role in ensuring the proper development of the embryo \cite{valet2022mechanical,petridou2019tissue} and as it progresses through different developmental stages, it is likely that its mechanical properties must also change over time. For example, cell elasticity can play a role in successful embryogenesis. Indeed, with optical-tweezers microrheology (see \cref{tbox:methods} and \cref{tbox:microrheo}), it was observed that zebrafish gut progenitors have a higher elasticity than their neighbouring cells. This increase generated by a microtubule dominant cytoskeleton is assumed to compensate for the dynamic remodeling of surrounding softer liver progenitor cells \cite{dzementsei2022foregut}. Furthermore, changes in cell elasticity are sufficient to drive lineage segregation during embryogenesis between epiblast and primitive endoderm within the mouse embryo \cite{ritter2022differential}. Additionally, cell rigidity acts as a protective mechanism, since a stiffening of the trophoblast and inner cell mass as well as an increase in tight junctions serves as a prevention mechanism for fluid loss through the epithelial layer during mouse blastocysts progression \cite{wang2018characterizing}. While augmentation in rigidity provides a shield against external forces and preserves structural integrity, softening is equally relevant. For instance, in blastoderm during early cellularization, it limits the extensive tissue deformations required for \textit{drosophila} morphogenesis \cite{d2019vivo}. Additionally, softening and fluidification of the cytoplasm was found to be a sign of ageing in \textit{C. elegans}, which is enhanced by nuclear envelope protein defects \cite{catala2025measuring}.

\subsection{Dynamical mechanical changes across the cell cycle}

The cell division cycle is another example in which mechanical properties vary over time, since there are changes in shape and volume observed \cite{clark2011mechanics}. As cells are preparing for mitosis, there is a remodelling of the cytoskeleton \cite{taubenberger2020mechanics} leading to a stiffening of the cell cortex \cite{fischer2016rheology}. Viscoelastic properties also fluctuate throughout the division cycle with an increased stiffness and lowered fluidity from G1 through the G1/S to the S/G2/M phases in melanoma cell lines \cite{schachtele2022combined}. Taking a closer look at mitosis, a softening of the cytoplasm as well as a decrease in intracellular activity reflects the reduction of actin polymerization or myosin activity, likely due to cortical recruitment of both components that is known to stiffen the cortex. This suggests that cells increase their global stiffness via cortical enforcement, thereby protecting the simultaneously softened intracellular space to ensure chromosome segregation while maintaining cellular integrity \cite{hurst2021intracellular}. Such conclusions can be drawn thanks to the combination of active and passive microrheology \cite{hurst2021intracellular, ahmed2015active}. Studies on cell division using only passive particle tracking should not be used to extract mechanical information \cite{chen2014intracellular}, without explicitly taking the active particle mobility into account \cite{muenker2024accessing}. This is important as only the combination of these measurements does not inherently assume that cells are in thermodynamical equilibrium, and hence essentially dead. Therefore, care must be taken when interpreting intracellular particle tracking to obtain mechanical information (see \cref{tbox:microrheo} for details). 

\subsection{Mechanical transitions during cellular differentiation}


In contrast to cell division, cellular differentiation occurs on longer timescales of days and weeks. As pluripotent cells differentiate into specialized cells, there are changes in shape, size and consequently in their internal structural organization, thus leading to potential variations in the mechanical properties over time. For instance, stem cell-derived adipocytes exhibit a continuous softening \cite{he2023changes}, a process linked to a decreased of F-actin levels \cite{chen2018actin}. This accommodation of the cytoskeleton allows lipid accumulation and ensures the proper metabolic function \cite{abuhattum2022adipose}. These findings obtained with AFM (see \cref{tbox:methods}) were supported by intracellular investigation employing video particle tracking microrheology (see \cref{tbox:methods} and \cref{tbox:microrheo}), revealing a consistent trend \cite{chen2016bio}. However, conducting these experiments on suspended cells, methods like micropipette aspiration (see \cref{tbox:methods}) have been used to observe a cell stiffening that is associated with cell spreading, before a softening sets in as the cells became fully committed \cite{yu2010mechanical}.

Conversely, the ongoing investigation of the mechanics during osteogenic differentiation led to very conflicting results. On one hand, the commitment to bone cells resulted in a drop in Young's modulus as derived from AFM measurements \cite{titushkin2007modulation, chen2010afm}. This softening is attributed to a remodelling of the cytoskeletal components, mainly the shift from thick actin stress fiber as known in mesenchemial stem cells to a thinner fiber network found in osteoblasts \cite{titushkin2007modulation, chen2010afm}. More recent experiments challenge these finding by reporting a stiffening during differentiation that is observed using AFM \cite{he2023changes, yen2020alteration} and micropipette aspiration \cite{yu2010mechanical}. To which extend this difference can be explained by potential changes in intracellular mechanics remains to be studied. 

More consistent are studies on smooth muscle cells and cardiomyocytes, both derived from the mesoderm, that present a stiffening during differentiation. These findings have been obtained by both micropipette aspiration  \cite{khani2014evaluation} and optical tweezers stretching \cite{tan2012probing}. Such consistent changes of mechanical properties suggests to be linked to the biological function of muscle cells, which have to withstand mechanical forces \cite{tan2012probing}. Finally, recent work highlighted the differences in intracellular stiffness and fluidity as well as cell activity across a variety of cell types, by a combination of active and passive (see \cref{tbox:microrheo}) optical tweezers experiments \cite{muenker2023intracellular}. Although this study was not conducted during differentiation, it still provides valuable insights on intracellular mechanics and paves the way to establish a better connection between cellular function and mechanics.

The variation in mechanical properties reported across differentiation studies may arise from differences in experimental conditions, such as cell origin \cite{chen2017mechanical}, cell substrate \cite{engler2006matrix, yen2020alteration}, cell shape (spread or spherical) \cite{darling2008viscoelastic} or even measurement techniques. For example, differences in Young's modulus between vascular smooth muscle cells derived from bone marrow and those from adipose tissue highlight the influence of cell source \cite{chen2017mechanical}. Moreover, discrepancies in AFM-based measurements can result from factors like tip geometry, which may overestimate elasticity \cite{harris2011experimental}. Despite these inconsistencies, examining the temporal evolution of intracellular and cellular mechanics across all the biological processes mentioned above allows valuable insights into fundamental biology, bridging the gap between physical properties and cellular behavior.

\section{Activity as important contribution to intracellular mechanics}
\label{sec:activity}
\pdfoutput=1

\subsection{Why activity matters}
While spatial and temporal dynamics of the cytoskeleton are important for changing the intracellular viscoelastic properties, there is increasing evidence that mechanics can also be altered quite drastically without actual structural changes, but by controlled variation of active forces. This complex interplay between cellular activity and cell mechanics is a fascinating, and emerging research field that uncovers how cells can rapidly react to mechanical challenges. When cells perform their various functions, they undergo a range of mechanical adaptations that are essential for their survival, growth, and division. As explained, the activity and force generation capacity of the cytoskeleton plays an important role here \cite{fletcher2010cell}.

First and foremost, an increase in active forces contributes to increased organelle movement within the cell. This internal transportation is an essential function that permits cells to distribute resources and get rid of waste \cite{brangwynne2009intracellular}. In bacteria it is known that an increased cellular metabolism does fluidize the cytoplasm, leading to different components of the cell to spread inside the cytoplasm \cite{parry2014bacterial}.

\subsubsection*{Active intracellular diffusion during cell division}

Mitosis is one of the most important biological events, which is intrinsically mechanical in nature. The cell cortex thickens and stiffens \cite{stewart2011hydrostatic}, potentially to protect the fragile intracellular processes taking places inside the cell. The chromosomes are pulled apart, and the two daughter cells get separated by a contractile actin ring. The intracellular mechanics not only changes by softening, but also by a change in effective energy spent to move particles \cite{hurst2021intracellular}. Effectively, the mobility of organelles increases \cite{carlini2020microtubules}, which has been suggested to be an efficient way of reaching an equal distribution of organelles, even in cells that had been highly polarized before mitosis setting in \cite{moore2021actin}. The activity is not only important for the regulated transport, but is elegantly used to exploit the enhanced active diffusion to ensure thorough mixing of organelles and cytoplasm during mitosis.

\subsubsection*{Dependence of activity on cross-linking proteins}
As described above, mechanical properties of the cytoskeleton are largely defined by local activation of cross-linking proteins, and the structure of cross-linking molecules determines many structural mechanical properties inside cells. While cross-linking proteins are typically passive, the actin based motor protein myosin represents a highly efficient cross-linker, that not only provides force generation, but also transmits stress across filaments. In fact, it largely depends on these cross-linkers whether activity in the intracellular space and in actin networks \textit{in vitro} leads to drastically different outcomes, i.e. fluidization or stiffening.

\subsection{Active fluidization vs. active cell stiffening}
Mastering this alternation between fluid-like and solid-like states is essential for cells to carry out many biological functions. Therefore, understanding what enables cells to rapidly switch between a dominantly fluid or rigid phenotype is an essential part of understanding its internal mechanics. With this dynamic interplay, activity can lead to stiffening of the cell on one hand and softening or fluidization on the other. The comparison between the cell cortex and the cytoplasm shows this activity dependent dichotomy.

It is well established for the cell cortex that increased activity results in a pulling force exerted on actin filaments, eventually resulting in stress stiffening \cite{cartagenarivera2016actomyosin}. This mechanism plays a role in maintaining the structural integrity of the cell and allows it to withstand external forces \cite{hu2019mechanical}. It was also shown \textit{in vitro} that stiffening of the cytoskeleton is sensitive to changes in motor activity. In fact, myosin motors can stiffen a reconstructed network of filaments by more than 100 fold simply via pulling on actin filaments \cite{koenderink2009active}. This stiffening can happen over several length scales within seconds when intracellular forces are applied. Such impressive changes in physical properties happen even without drastic structural changes, but can be simply triggered by intracellular signalling of myosin contractility. These results emphasize the rapid adaptation of the cytoplasm to mechanical forces acting on the interior of a cell and the associated possibility that activity can lead to stiffening of the cytoskeleton \cite{tang2024stiffening}.

In contrast, exactly the same signalling of myosin contractility can also lead to the opposite outcome: Fluidization. The shift from a stiff to a more fluid state can occur due to filaments gliding past each other, enabling the cell to adapt to its changing needs or environmental conditions.
This fluidization happens not at the cortex, but at the intracellular space and has been observed to be triggered by activity. The transition to softer and more fluid characteristics resembles increased active diffusion which is powered by complex assemblies of molecular motors acting on the cytoskeletal structures. For example in \textit{Drosophila} oocytes, such active diffusion is driven by the actomyosin cytoskeleton \cite{drechsler2017active}. This process has also been investigated for vesicles in mouse oocytes, where it was shown that gradients of active diffusion can position the nucleus during oocyte maturation. Additionally, the motor activity renders the intracellular region of these oocytes softer, effectively contradicting the stress stiffening observed in the cell cortex. These differences suggest that the structural connectivity between the actin fibers and the molecular motors are highly relevant to control the final outcome of active forces
\cite{almonacid2015active,ahmed2018active}.
Besides mammalian cells, similar active fluidization has been shown in the plant cytoskeleton. Here, the inhibition of myosin activity in the presence of actin filaments was demonstrated to stiffen the tissue and that myosin activity leads to increased movement of the filaments \cite{honing2009actin}.

The explanation for these contrary effects of active force generation on seemingly similar structures can be derived theoretically, and the contradiction is quickly resolved when taking the cross-linkers into account. For cross-linkers where activity favors linker unbinding, an increase in active forces leads to a decrease in viscosity, which eventually triggers the fluidization of the material \cite{oriola2017fluidization}. In the extreme case where cross-linkers are absent, the active contractile forces generated by motor proteins lead to an increase of cytoskeletal filament sliding, thus quickly resolving any elasticity that is maintained by filament entanglement \cite{fuerthauer2019self-straining}. Conversely, if the cross-linkers continue to link the filaments even under load, the system stiffens, mainly because filament bending decreases and is replaced by the stretching and compression of the filaments, which generates much larger stiffness at the material level. 

A striking, and up to now not understood fact is that in detached cells in suspension, the activity rather results in a fluidization, whereas in cells that are attached to a substrate, motor activity stiffens the cell \cite{chan2015myosin}.

\section{Summary and open questions}
\label{sec:summary}
\pdfoutput=1

While the importance of forces and viscoelastic properties at the cell and tissue levels is well established \cite{phuyal2023mechanobiology, vogel2018unraveling, goodwin2021mechanics,pavin2021mechanobiology}, we are only beginning to understand the implications of intracellular active mechanics for cell biology. In the light of findings from the past decade, it has become clear that three key contributions of intracellular mechanics to cell biological functions must be considered when trying to understand the functioning of living cells: Heterogeneity, reaction rates, and activity; each require more attention for further research.

First, the effect of spatial and temporal heterogeneity in viscoelastic properties is well documented, with ample evidence supporting its functional relevance. Simply put, the cytosol should not be viewed as a homogeneous soup of proteins, molecules, and filaments. Instead, it has the capacity to locally solidify or liquefy, enabling it to push organelles and large macromolecules to new positions or retain them despite active and passive diffusion. This heterogeneity underpins the spatial organization within cells and directly affects intracellular transport and dynamics. But how is this heterogeneity established, and how is it maintained? Does it simply stabilize the cell, or directly interact with biochemical reactions? These are important questions to be addressed in the future.

Second, the fact that molecular and organelle mobility, as well as viscoelastic properties, vary over several orders of magnitude suggests that many chemical reactions are at least partially diffusion-limited. This has profound implications for biochemical reaction networks, which are inherently nonlinear in nature. For instance, waves of signaling molecules such as calcium or small RhoGTPases are typically explained by reaction-diffusion systems. A global or local change in either reaction rates or diffusion coefficients can drastically alter signaling dynamics, potentially halting oscillations entirely. Such effects are direct outcomes of high Hill coefficients, which reflect the mathematical nonlinearity of these systems. Consequently, biological processes like quorum sensing, signaling waves, and molecular switches---known to be critical for cellular function---are highly sensitive to changes in viscoelastic properties and mobility. But which reactions are affected by diffusion? And is this influence perhaps the key to understanding the regulation of intracellular mechanics?

Third, active forces, which added to thermal motion, play a significant role. In recent years, a series of independent experiments has demonstrated that random active forces significantly increase organelle and probe particle mobility. While this may initially appear to be an energetically wasteful process, these active forces have profound effects on effective diffusion. In highly crowded intracellular environments, where thermal (Brownian) motion is significantly reduced, active forces can rescue transport and stabilize biochemical reactions. Moreover, these active forces can lead to drastic changes in the viscoelastic properties of the cytoplasm, such as active fluidization or, conversely, active stiffening. Yet, to which extent these active forces are indeed important for cellular function remains to be demonstrated.

Given that most, if not all, biological functions depend on intracellular organization and biochemical reactions, it is evident that intracellular active mechanics must be precisely regulated by the cell. This raises a series of critical questions for future research: What are the physical boundaries constraining intracellular mechanics? How do cells locally and temporally regulate these properties? Emerging evidence suggests that precise control of protein density is involved, but how cells measure local molecular density and crowding remains unknown. An exciting challenge for the near future is to identify and understand the proteins responsible for sensing and regulating intracellular mechanics. Promising candidates include intrinsically disordered proteins and elastic molecular domains, which are particularly susceptible to active random forces.

The next decade promises significant advances in the quantification and functional characterization of intracellular active mechanics. These efforts will provide a more holistic view of this key parameter, which has been largely neglected so far.

\section*{Acknowledgements}
The authors would like to thanks Till Muenker for providing a blender model to generate the figures of this review. This work was supported by the Deutsche Forschungsgemeinschaft (DFG, German Research Foundation): Project-ID 449750155 – RTG 2756, Projects B3, Project-ID 516046415 and under Germany’s Excellence Strategy (EXC 2067/1- 390729940).

\printbibliography

\end{document}